\documentclass[10pt]{blois07}
\usepackage{graphicx,amsmath,amssymb,cite}
\newcommand{\Pom}{{\mathbb{P}}}

\begin{document}
\title{Diffractive parton distributions:\\
the role of the perturbative Pomeron}
\author{G.~Watt$^1$, A.D.~Martin$^2$ and \underline{M.G.~Ryskin}$^3$
\protect\footnote{\ \ Talk presented at the 12th 
International Conference on Elastic and Diffractive Scattering: Forward 
Physics and QCD, DESY, Hamburg, Germany, 21-25 May 2007}}
\institute{
$^1$ Department of Physics \& Astronomy, University College London, 
WC1E 6BT, UK\\
$^2$ Institute for Particle Physics Phenomenology, University of Durham, 
DH1 3LE, UK\\
$^3$ Petersburg Nuclear Physics Institute, Gatchina, St.~Petersburg, 
188300, Russia
}

\maketitle

\begin{abstract}
  We consider the role of the perturbative Pomeron-to-parton splitting in the 
  formation of the diffractive parton distributions.
\end{abstract}

\begin{figure}[t]
\centering
(a)\hspace{0.3\textwidth}(b)\hspace{0.3\textwidth}(c)\\
\begin{minipage}{0.35\textwidth}
  \includegraphics[width=\textwidth]{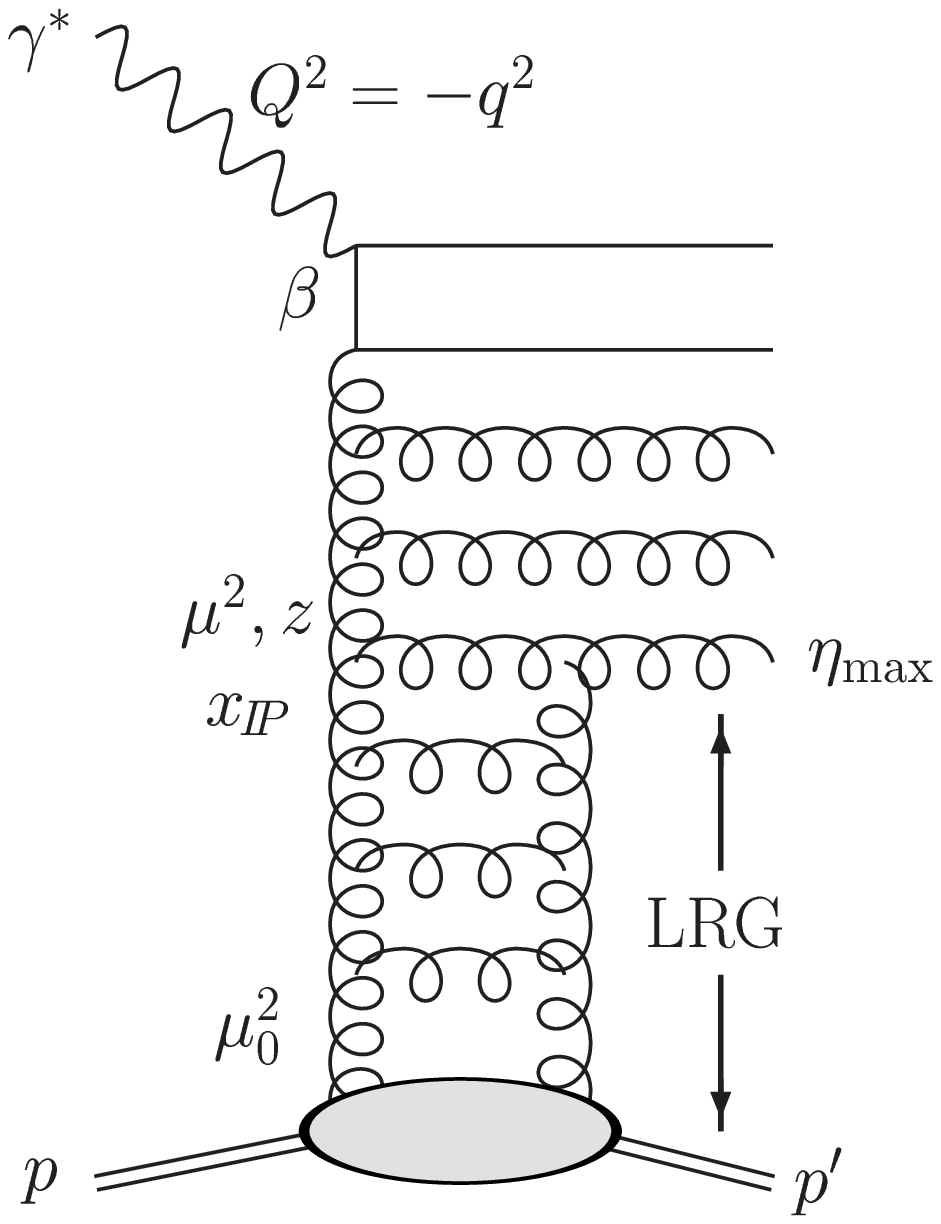}
\end{minipage}
\begin{minipage}{0.3\textwidth}
  \includegraphics[width=\textwidth]{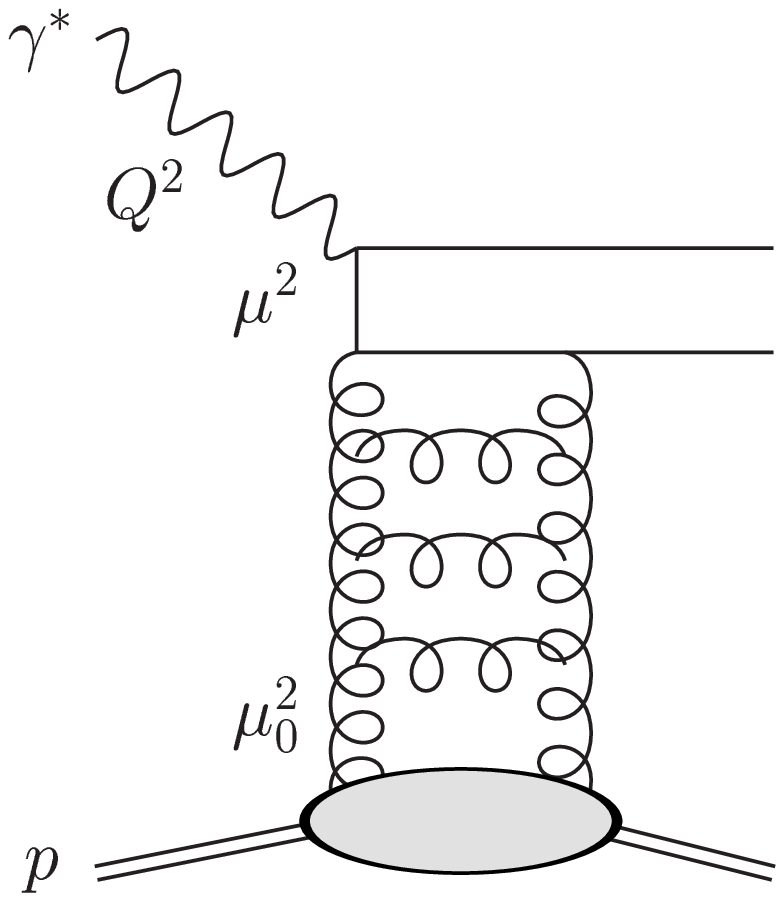}
\end{minipage}\hfill
\begin{minipage}{0.3\textwidth}
  \includegraphics[width=0.8\textwidth]{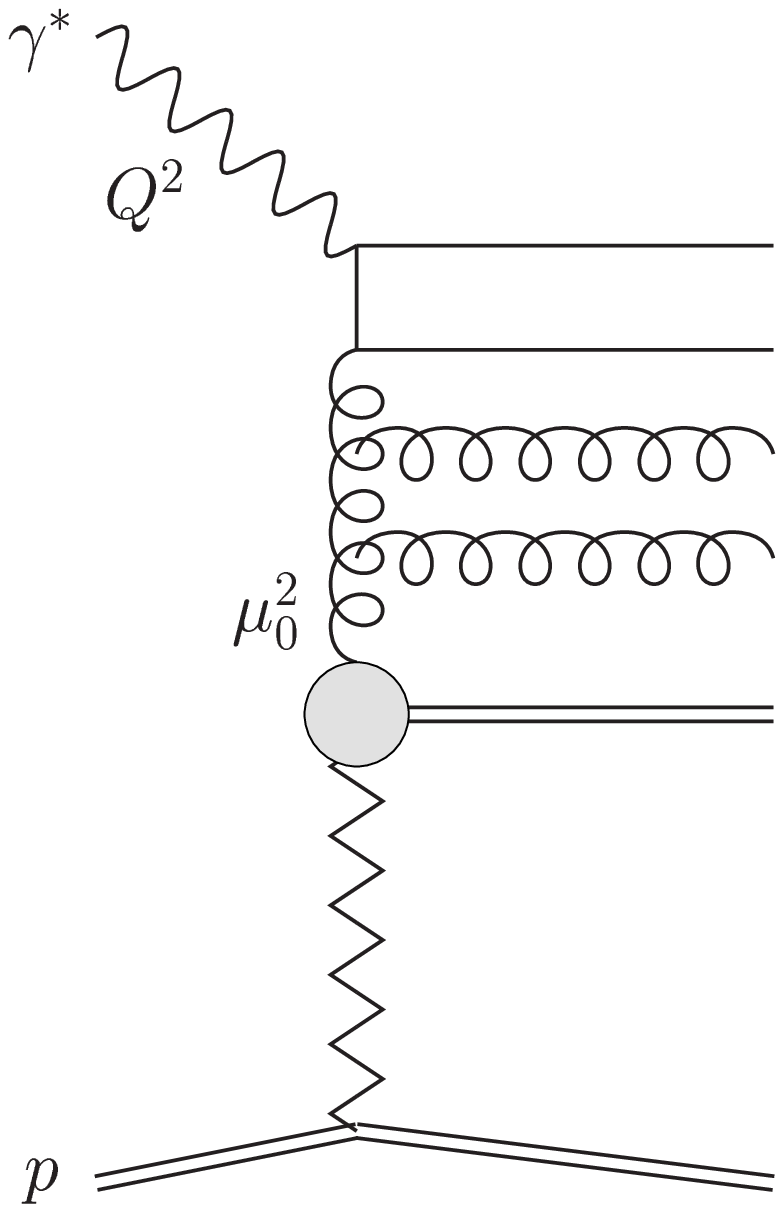}
\end{minipage}
\caption{(a) The perturbative \emph{resolved} Pomeron contribution, which is 
the basis of the perturbative QCD approach, (b) the perturbative \emph{direct}
 Pomeron contribution, and (c) the non-perturbative \emph{resolved} Pomeron 
diagram, which accounts for the contribution from low scales, $\mu <\mu_0$.}
\label{fig:graph}
\end{figure}

Diffractive deep-inelastic scattering (DDIS), $\gamma^*p\to X+p$, is
characterised by a large rapidity gap (LRG) between the cluster $X$ of
outgoing hadrons and the slightly deflected proton, understood to be
due to `Pomeron' exchange.  Let the momenta of the incoming proton, the
outgoing proton, and the photon be labelled $p$, $p^\prime$ and $q$
respectively; see Fig.~\ref{fig:graph}(a).  Then the basic kinematic
variables in DDIS are the photon virtuality, $Q^2=-q^2$, the Bjorken-$x$
variable, $x_B=Q^2/(2p\cdot q)$, the squared momentum transfer, $t=(p-p')^2$,
the fraction of the proton's light-cone momentum transferred through the LRG,
$x_{\Pom}=1-{p^\prime}^+/p^+$, and $\beta\equiv x_B/x_{\Pom}$.

It is common to perform analyses of DDIS data based on two levels of
factorisation.  First the $t$-integrated diffractive structure function
$F_2^{{\rm D}(3)}$ may be written as the convolution of the usual coefficient
functions $C_{2,a}$ as in DIS with diffractive parton distribution functions
(DPDFs)
$a^{\rm D}$ \cite{Collins:1997sr}:
\begin{equation} \label{eq:collfact}
  F_2^{{\rm D}(3)}(x_{\Pom},\beta,Q^2)\ =\ 
\sum_{a=q,g}\;\beta\int_\beta^1\!\frac{{\rm d}z}{z^2}\;C_{2,a}
\left(\frac{\beta}{z}\right)\;a^{\rm D}(x_{\Pom},z,\mu_F^2),
\end{equation}
with the factorisation scale $\mu_F$ usually taken to be $Q$.  The DPDFs
$a^{\rm D}=z q^{\rm D}$ or $z g^{\rm D}$ satisfy DGLAP evolution in 
$\mu_F$.  The convolution variable $z\in[\beta,1]$ is the
 fraction of the Pomeron's light-cone momentum carried by the 
parton entering the hard subprocess.
 At leading-order (LO) the coefficient functions are 
$C_{2,q}(x)=e_q^2\,\delta(1-x)$ and $C_{2,g}(x)=0$.  The collinear
factorisation theorem \eqref{eq:collfact} applies when $\mu_F$ is made very
large; it is correct up to power-suppressed corrections.  In the
second stage, Regge factorisation is usually assumed, such that the
diffractive parton densities $a^{\rm D}$ are written as the product 
of the Pomeron flux factor
$f_{\Pom}=\int\!{\rm d}t\;\exp(B_{\Pom}\,t)\;x_{\Pom}^{1-2\alpha_{\Pom}(t)}$
and the Pomeron parton densities $a^{\Pom}=z q^{\Pom}$ or $z g^{\Pom}$,
that is, \begin{equation} \label{eq:reggefact}
a^{\rm D}(x_{\Pom},z,\mu_F^2)=f_{\Pom}(x_{\Pom})\,a^{\Pom}(z,\mu_F^2).
\end{equation}
The Pomeron trajectory is
$\alpha_{\Pom}(t)=\alpha_{\Pom}(0)+\alpha'_{\Pom}\,t$.  For simplicity, we
omit the contribution of secondary Reggeons to the right-hand side of
\eqref{eq:reggefact}.  Such an approach says nothing about the mechanism
for diffraction: information about the diffractive exchange (`Pomeron')
needs to be parameterised at the input scale $\mu_0$ and fitted to the data.

An alternative way to describe DDIS is to consider the heavy photon
transition to $q\bar{q}$ or effective $(q\bar{q})g$ dipoles,
which then interact with the target proton via two-gluon
exchange \cite{Wusthoff:1997fz,GolecBiernat:1999qd}; the case of the 
$q\bar{q}$ dipole is shown in Fig.~\ref{fig:graph}(b).  Here the Pomeron
is modelled by the $t$-channel colour-singlet pair of gluons and we
have an explicit form of the Pomeron parton distributions $a^{\Pom}(z,\mu^2)$
given at the initial virtuality $\mu^2=k^2_t/(1-z)$ fixed by the transverse
momentum $k_t$ of the outgoing components of the dipole.  Such an approach 
relies on the existence of a large saturation scale 
$Q_S^2\gtrsim 1$ GeV$^2$ \cite{GolecBiernat:1998js} to act as 
an infrared cutoff and suppress the contribution from large dipole 
sizes, thereby justifying the use of perturbative QCD for the whole 
$F_2^{{\rm D}(3)}$.  However, more sophisticated dipole models 
\cite{Kowalski:2006hc} now give a lower $Q_S^2\lesssim 0.5$ GeV$^2$ 
in the HERA kinematic regime, therefore significant non-perturbative 
contributions to inclusive DDIS are to be expected.

The two theoretical frameworks are essentially contradictory: the
Regge factorisation approach is motivated by `soft' Pomeron exchange,
whereas the two-gluon exchange approach is motivated by perturbative QCD. 
Nevertheless, a `doublethink' mentality exists whereby the two approaches
are both commonly applied separately in the description of data, often
within the same paper, with few attempts made to reconcile them.

The two approaches can be combined
\cite{Martin:2004xw,Martin:2005hd,Martin:2006td} by 
generalising the $\gamma^*\to q\bar{q}$ and $\gamma^*\to q\bar{q}g$ 
transitions to an arbitrary number of parton emissions in the final 
state, as shown in the upper half of Fig.~\ref{fig:graph}(a).  The 
perturbative Pomeron is described by a parton ladder ending in a pair 
of $t$-channel gluons or sea-quarks; the former is shown
in the lower half of Fig.~\ref{fig:graph}(a).  The virtualities of 
the $t$-channel partons are strongly ordered as required by DGLAP evolution: 
$\mu^2_0\ll\ldots\ll\mu^2\ll\ldots\ll\mu_F^2$.  The scale $\mu^2$ at 
which the Pomeron-to-parton splitting occurs can vary between 
$\mu^2_0\sim1$ GeV$^2$ and $\mu_F^2$.  For $\mu<\mu_0$, the representation 
of the Pomeron as a perturbative parton ladder is no longer valid, and 
instead, in the lack of a precise theory of non-perturbative QCD, we appeal
to Regge theory where the `soft' Pomeron is treated as an effective Regge pole 
with intercept $\alpha_{\Pom}(0)\simeq1.08$; see Fig.~\ref{fig:graph}(c).

The probability to find an appropriate pair of $t$-channel gluons with 
transverse momentum $l_t$, integrated over $l_t$, is given by the usual 
gluon distribution of the proton obtained from the global parton 
analyses, $x_{\Pom}g(x_{\Pom},\mu^2)$, where $\mu^2=k_t^2/(1-z)$ is the 
virtuality of the first $t$-channel parton in the upper part of the 
diagram.  The emitted parton at the edge of the LRG in Fig.~\ref{fig:graph}(a)
 has rapidity $\eta_{\rm max}$ and transverse momentum $k_t$ and carries a 
fraction $(1-z)$ of the Pomeron's light-cone momentum.  For inclusive DDIS 
we must integrate over $k_t$, accounting for the components of the Pomeron 
wave function of different sizes $\sim1/k_t$, which translates to an integral
 over $\mu$ of the form
\begin{equation} \label{eq:adpert}
  a^{\rm D}(x_{\Pom},z,\mu_F^2)\ =\ \int\limits^{\mu_F^2}_{\mu^2_0}
\frac{{\rm d}\mu^2}{\mu^2}\;\frac{1}{x_{\Pom}}
  \left[R_g\frac{\alpha_S(\mu^2)}\mu x_{\Pom}g(x_{\Pom},\mu^2)\right]^2\;
a^{\Pom}(z,\mu_F^2;\mu^2).
\end{equation}
The term $f_\Pom(x_\Pom;\mu^2)\equiv [\ldots]^2/x_{\Pom}$ plays the role
of the Pomeron flux in 
\eqref{eq:reggefact}.  $R_g$ is the skewed factor which accounts for the 
fact that in the lower parts of Fig.~\ref{fig:graph}(a,b) we deal not 
with the diagonal but with the skewed (or generalised) parton distributions.
  For low values of $x_{\Pom}\ll 1$ this skewed factor is given by the
Shuvaev prescription \cite{Shuvaev:1999ce}.  The notation
$a^{\Pom}(z,\mu_F^2;\mu^2)$ for the Pomeron parton densities means
that they are DGLAP-evolved from 
an initial scale $\mu^2$ up to the factorisation scale $\mu_F^2$.

At first sight, the integral \eqref{eq:adpert} appears to be concentrated
in the 
infrared region of low $\mu$.  However, for DDIS we consider very small 
$x_{\Pom}$ values.  In this domain, the gluon distribution of the proton 
has a large anomalous dimension.  Asymptotically, as $x_{\Pom}\to 0$, 
BFKL predicts $x_{\Pom}g(x_{\Pom},\mu^2)\sim (\mu^2)^{0.5}$ for fixed 
$\alpha_S$ \cite{Lipatov:1996ts}.  In this case the integral 
\eqref{eq:adpert} takes the logarithmic form, and we cannot neglect 
the contribution from large scales $\mu$.

By differentiating \eqref{eq:adpert} with respect to $\ln(\mu_F^2)$, 
the evolution equation now reads \cite{Martin:2005hd}
\begin{equation} \label{eq:adevolve}
  \frac{\partial a^{\rm D}(x_{\Pom},z,\mu^2)}
       {\partial\ln\mu^2}\;=\; \sum_{a^\prime=q,g}P_{a,a^\prime}\otimes 
       a^{\prime\,{\rm D}}\ +
       \sum_{\Pom=G,S,GS}P_{a,\Pom}(z)f_{\Pom}(x_{\Pom};\mu^2)\,.
 \end{equation}
Here the first term, involving the usual parton-to-parton splitting 
functions, $P_{a,a^\prime}$, arises from DGLAP evolution in the upper parts 
of Fig.~\ref{fig:graph}(a,c).  The second (inhomogeneous) term, 
involving the Pomeron-to-parton splitting functions, 
$P_{a,\Pom}(z)\equiv a^{\Pom}(z,\mu^2;\mu^2)$, arises from the transition 
from the two $t$-channel partons (that is, the Pomeron) to a single 
$t$-channel parton in Fig.~\ref{fig:graph}(a).  The notation $\Pom=G,S,GS$ 
denotes whether the uppermost two t-channel partons are gluons ($\Pom=G$) or 
sea-quarks ($\Pom=S$)\footnote{Note that besides the two-gluon exchange
we also account for the
$q\bar{q}$ $t$-channel colour-singlet state, since even at very
small $x_{\Pom}$ the sea-quark contribution does not die out.  Moreover,
the skewed factor $R_q$ for the $q\bar{q}$ $t$-channel state is more than
three times larger than that for the gluons \cite{Shuvaev:1999ce}.}
, while the interference term is denoted by $\Pom=GS$. 
The LO Pomeron-to-parton splitting functions, $P_{a,\Pom}$, were calculated 
in \cite{Martin:2005hd}, where the perturbative Pomeron flux factors, 
$f_{\Pom}(x_{\Pom};\mu^2)$, are also given.

Simultaneously, we need to add in \eqref{eq:collfact} the direct 
Pomeron--photon interaction described by the coefficient function 
$C_{2,\Pom}=C_{T,\Pom}+C_{L,\Pom}$ corresponding to the hard 
subprocess shown in Fig.~\ref{fig:graph}(b):
\begin{equation} \label{eq:f2d3}
  F^{{\rm D}(3)}_2\ =\ \sum_{a=q,g}C_{2,a}\otimes a^{\rm D}\; +
  \sum_{\Pom=G,S,GS}C_{2,\Pom}.
\end{equation}
For the LO $C_{T,\Pom}$ the light-quark contributions in the limit 
$\mu^2\ll Q^2$ are subtracted since they are already included in the 
first term of \eqref{eq:f2d3} via the inhomogeneous evolution of the DPDFs.  
This subtraction defines a choice of factorisation scheme.  There is no
such subtraction for the LO $C_{L,\Pom}$, which are purely higher-twist,
or for the heavy quark contributions since we work in the fixed
flavour-number scheme (FFNS).

Thus, we see from \eqref{eq:adevolve} and \eqref{eq:f2d3} that the 
diffractive structure function is analogous to the photon structure 
function, where there are both resolved and direct components and 
where the photon PDFs also satisfy an
inhomogeneous evolution equation.

As usual the evolution starts from some not-too-small scale $\mu_0$
and all the contributions, both perturbative and non-perturbative,
coming from $\mu\le\mu_0$ are parameterised in terms of the Regge
factorisation as some input which should be fitted to the data.
Note that inclusion of the inhomogeneous term in \eqref{eq:adevolve} and
the direct Pomeron coefficient function in \eqref{eq:f2d3} does not
add any new free parameters to the description of the DPDFs.  The LO
Pomeron-to-parton splitting functions are known \cite{Martin:2005hd}
and at next-to-leading order (NLO) they can be calculated unambiguously.  
The numerical results \cite{Martin:2006td} presented below were obtained by 
fitting the H1 LRG data \cite{Aktas:2006hy} with $M_X>2$ GeV and $Q^2\ge 8.5$ 
GeV$^2$, adding statistical and systematic experimental errors in quadrature, 
and taking the input distributions at $\mu_0^2 = 2$ GeV$^2$, which gave a 
$\chi^2/\mathrm{d.o.f.}$ of $0.84$.
\begin{figure}[t]
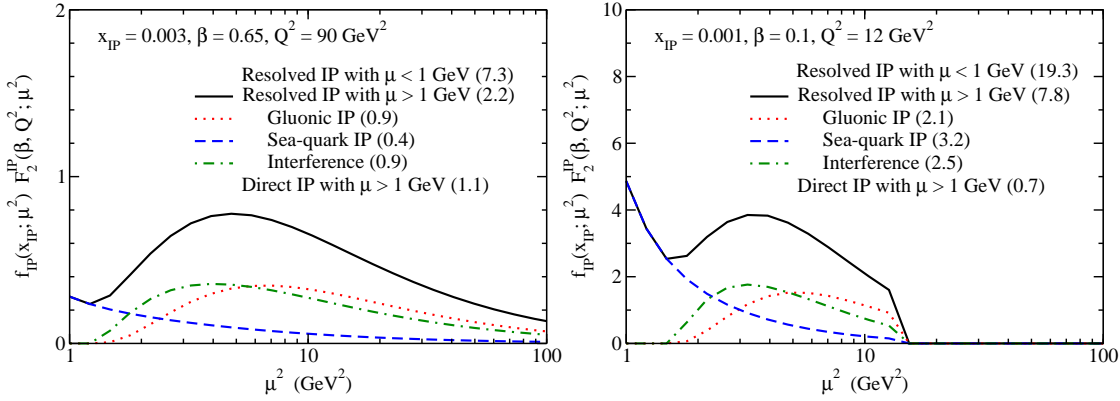

\centering
  \includegraphics[width=0.5\textwidth,clip]{f2d3pert90.eps}%
  \includegraphics[width=0.5\textwidth,clip]{f2d3pert12.eps}
\caption{The breakdown of the resolved Pomeron contribution for $\mu>1$ 
GeV to the total $F^{{\rm D}(3)}_2$ as a function of $\mu^2$ for two 
representative values of $(x_{\Pom},\beta,Q^2)$.  The integral over 
$\ln(\mu^2)$ is shown by the numbers in parentheses in the legend.  Also 
shown are the integrated contributions from the the resolved Pomeron 
with $\mu<1$ GeV and the direct Pomeron with $\mu>1$ GeV.  The secondary 
Reggeon contributions are negligible for the values of $x_{\Pom}$ chosen 
here.}
\label{fig:break}
\end{figure}

From Fig.~\ref{fig:break} we see that integrating the
perturbative contributions to $F^{{\rm D}(3)}_2$ starting from $\mu=1\,$GeV
we collect up to a third of the whole diffractive structure function
$F^{{\rm D}(3)}_2$.  Of course, the numerous corrections (higher order
$\alpha_S$ corrections, power corrections, etc.) are not negligible at
such low scales as $\mu\sim1\,$GeV.  Nevertheless, this fact indicates
that an important part of the diffractive parton densities comes from
the relatively small size components of the `Pomeron'.  That is why
 fitting the DDIS data in terms of the Regge factorisation one needs
 the effective Pomeron intercept $\alpha_{\Pom}(0)\simeq1.12$ larger than that
 ($\alpha_{\Pom}(0)\simeq 1.08$) measured in the purely soft hadron--hadron
 scattering.

\begin{figure}[t]
\centering
\includegraphics[width=0.8\textwidth,clip]{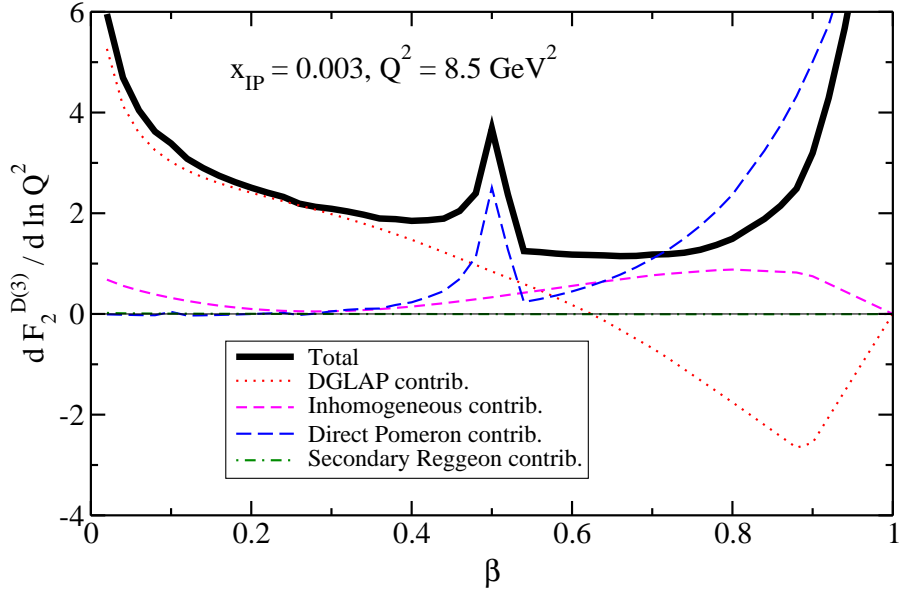}
\caption{The breakdown of the contributions to the slope
  $\partial F^{{\rm D}(3)}_2/\partial\ln Q^2$ as a function of $\beta$.  
Contributions apart from the usual DGLAP terms start becoming important for
$\beta\gtrsim 0.4$.  The peak at $\beta=Q^2/(Q^2+4m_c^2)\simeq 0.51$ is 
due to the threshold for charm production from the direct Pomeron--photon 
interaction.}
\label{fig:qsqslope}
\end{figure}
\begin{figure}[t]
  \centering
    \includegraphics[width=\textwidth,clip]{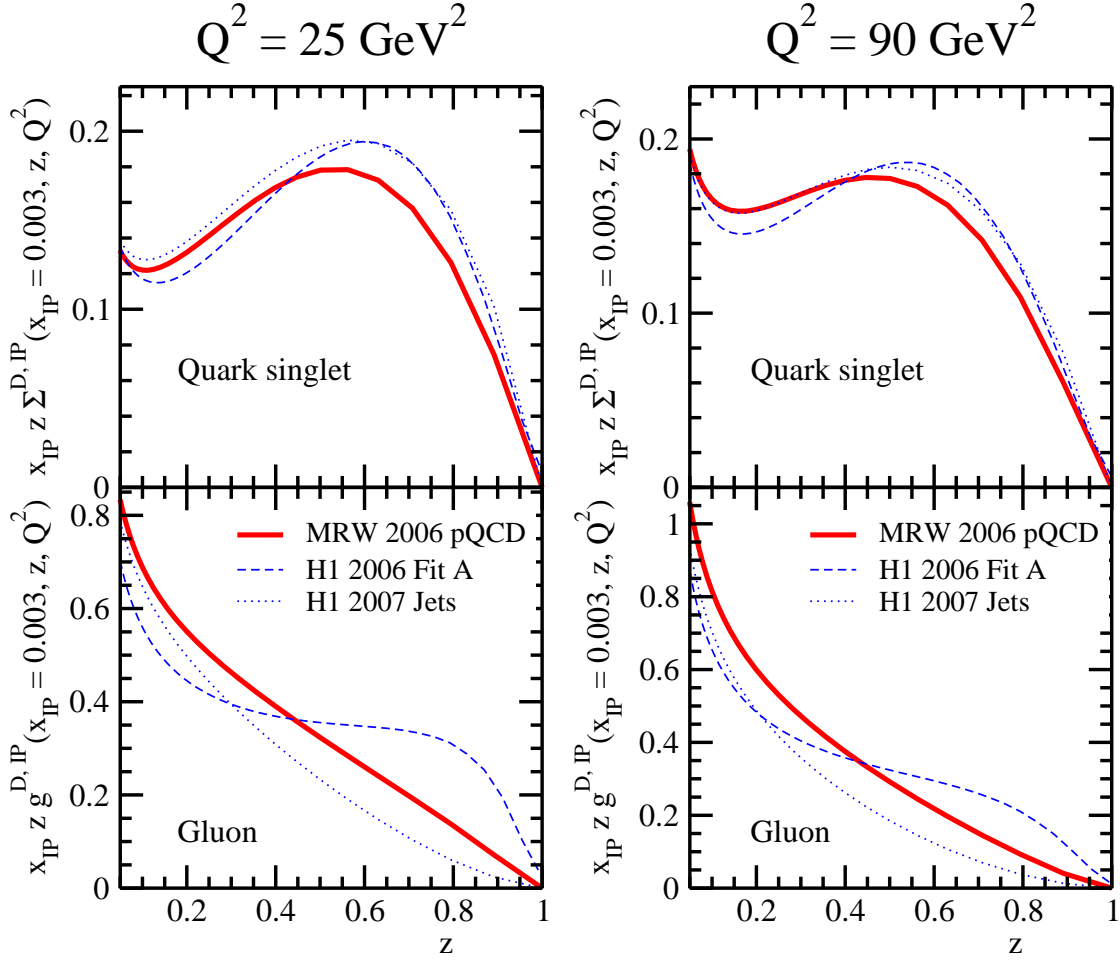}
    \caption{The MRW 2006 pQCD DPDFs compared to those from the H1 2006 
Fit A and the H1 2007 Jets fit.  All the DPDFs shown here are normalised to 
$M_Y<1.6$ GeV by multiplying by a factor 1.23 relative to $M_Y=m_p$, where 
$M_Y$ is the mass of the proton dissociative system.}
    \label{fig:fit}
\end{figure}

In Fig.~\ref{fig:qsqslope} we show the breakdown of the different
contributions 
to the slope $\partial F^{{\rm D}(3)}_2/\partial\ln Q^2$.  As it is seen, 
the inhomogeneous term in the evolution equation is fairly large and starts 
becoming important for $\beta\gtrsim 0.4$.  Note that the direct Pomeron
contribution at large $\beta$, 
mostly the twist-four $F^{{\rm D}(3)}_L$, only gives an important contribution 
to the total $F^{{\rm D}(3)}_2$ for $\beta\gtrsim 0.9$, but the contribution 
to the $Q^2$ slope starts becoming important at moderate $\beta$.  Therefore, 
it is not possible to avoid the presence of this contribution by simply 
excluding data points with $\beta>0.8$ from the fit, as is done in the H1 2006
 analysis \cite{Aktas:2006hy}.  In the analysis of inclusive DDIS data,
the diffractive gluon density is mainly determined by the derivative 
$\partial F^{{\rm D}(3)}_2/\partial\ln Q^2$.  The presence of these
additional positive 
contributions to the $Q^2$ slope at large $\beta$ apart from the usual 
DGLAP contribution means that a smaller gluon density is required for 
$z\gtrsim 0.4$ compared to the H1 2006 Fit A performed in the usual 
Regge factorisation framework \cite{Aktas:2006hy}; see Fig.~\ref{fig:fit}.

In our analysis we account for the twist-four
$F^{{\rm D}(3)}_L$ contribution coming from the direct Pomeron--photon 
fusion, that is, from the last term in \eqref{eq:f2d3} corresponding 
to Fig.~\ref{fig:graph}(b).  This contribution, which goes to a 
constant value as $\beta\to1$, was calculated in
\cite{Wusthoff:1997fz,GolecBiernat:1999qd} 
and turns out to be numerically appreciable.  For the coupling to massless 
quarks, this contribution to \eqref{eq:f2d3} takes the form 
\cite{Martin:2005hd}
\begin{equation} \label{eq:fld3}
  C_{L,\Pom} = \int_{\mu_0^2}^{\frac{Q^2}{4\beta}}\!{\rm d}\mu^2\;
  \frac{1/Q^2}{\sqrt{1-4\beta\mu^2/Q^2}}\;f_{\Pom}(x_{\Pom};\mu^2)\;
  F_L^{\Pom}(\beta).
\end{equation}
Differentiating with respect to $\ln(Q^2)$ gives a positive term from the 
integrand evaluated at $\mu^2=Q^2/(4\beta)$, analogous to the inhomogeneous 
term in the evolution equation, in addition to the negative term proportional 
to $-1/Q^2$ obtained from taking the derivative of the integrand.

A recent analysis \cite{GolecBiernat:2007kv} includes the twist-four 
$F_L^{{\rm D}(3)}$ contribution from all $k_t$ in addition to the usual 
Regge factorisation formulae of \eqref{eq:collfact} and 
\eqref{eq:reggefact}.  It is found that the twist-four $F_L^{{\rm D}(3)}$ term
gives a negative contribution to the $Q^2$ slope, such that a larger gluon
distribution at high $z$ is needed than in the usual Regge factorisation
approach.  However, the contribution from $\mu^2=k_t^2/(1-\beta)<\mu_0^2$ is 
already included in the input Pomeron PDFs taken at a scale
$\mu_0^2=1.5$ GeV$^2$, so some double-counting is involved.  Moreover, the DGLAP
evolution is not accounted for in the dipole cross section, and the
inhomogeneous term in the evolution equation of the DPDFs is neglected.  The 
future measurement of 
$F^{{\rm D}(3)}_L$ \cite{Newman:2005mm} will provide an important check 
of the calculations of the twist-four component.

The smaller diffractive gluon density at high $z$ found in our analysis, 
compared to the H1 2006 Fit A, is preferred by the data on inclusive 
diffractive dijet production, $\gamma^*p\to {\mbox{dijet}}\,X^\prime+p$, 
where at LO the dijet system originates from the outgoing $q\bar{q}$ pair 
in Fig.~\ref{fig:graph}(a,c) and the rest of the hadronic system $X^\prime$ 
originates from the other outgoing partons in Fig.~\ref{fig:graph}(a,c).  
However, the data still tend to be slightly overestimated \cite{Chekanov:2007yw}.  
This can also be seen in Fig.~\ref{fig:fit} where the `MRW 2006' gluon at high $z$ 
is smaller than the `H1 2006 Fit A' gluon, but larger than the `H1 2007 Jets' 
gluon obtained from a combined fit, 
within the Regge factorisation framework, to inclusive DDIS and inclusive 
diffractive dijet data \cite{H1jets}.  The $\chi^2$ for the 190 inclusive DDIS 
points increases from 158 (H1 2006 Fit A) to 169 (H1 2007 Jets fit) on 
inclusion of the dijet data.  Therefore, the gluon density determined 
indirectly from the inclusive DDIS data, under the assumption of pure DGLAP 
evolution, is different from the gluon density preferred by the dijet data.  
The apparent tension between the inclusive DDIS and dijet data in the Regge 
factorisation approach is partly alleviated by the inclusion of the 
perturbative Pomeron terms.

Note that in the `MRW 2006' analysis \cite{Martin:2006td} the parton 
distributions of the proton, representing the lower parton ladders in 
Fig.~\ref{fig:graph}(a,b), were taken from a NLO global fit.  However, since 
the Pomeron-to-parton splitting is only calculated at LO, it may be more 
appropriate to take the inclusive parton distributions from a LO fit, 
where the gluon density is much larger at small $x$ and does not take a 
valence-like form at low scales.  In this case, the inhomogeneous term in 
the evolution equation, and the other direct Pomeron contributions, would 
be enhanced, leading to an even smaller diffractive gluon density at 
high $z$.  Note that even if LO DGLAP evolution is used in the lower parton 
ladders of 
Fig.~\ref{fig:graph}(a,b), NLO DGLAP evolution may still be used for the 
evolution of the Pomeron PDFs in the upper parts of Fig.~\ref{fig:graph}(a,c).

A more direct way to observe the perturbative Pomeron contribution to
DDIS is to study the transverse momenta of secondaries in the `Pomeron
fragmentation' region, at the edge of the LRG, as in 
Ref.~\cite{Adloff:2000qi}.  
In contrast with the Regge
factorisation of Fig.~\ref{fig:graph}(c), where the
transverse momentum distribution of the partons inside a Reggeon is
assumed to have an exponential form with a rather low mean value of
intrinsic $k_t$, in the perturbative case of Fig.~\ref{fig:graph}(a) the
$k_t$-distribution of the lowest jet in Fig.\ref{fig:graph}(a) obeys a power
law, given by the integrand of \eqref{eq:adpert}.  Therefore we expect a
larger $k_t$ of the secondaries with the long power-like tail.

So far, measurements of inclusive diffractive dijets at HERA have primarily 
been made in the kinematic region of small $\beta\equiv x_B/x_\Pom$ where the 
resolved Pomeron mechanism of Fig.~\ref{fig:graph}(a,c) gives the dominant 
contribution, and the contribution from exclusive diffractive dijets, 
Fig.~\ref{fig:graph}(b), is negligible.  However, a first measurement has 
been made of dijets in DDIS with a cut on $\beta>0.45$ in order to enhance 
the contribution from exclusive diffractive dijets, 
$\gamma^{*}p\to\mbox{dijet}+p$ \cite{Szuba:2005dy}.  Within the HERA 
kinematic domain the sea-quark component of the Pomeron is quite important 
(see Fig.~\ref{fig:break}).  This statement can be
  checked by observing the diffractive high-$E_T$ dijet distributions
in the exclusive $\gamma^{(*)}p\to\mbox{dijet}+p$ process.  At LO we expect 
the
 ratio of cross sections for high-$E_T$ dijets produced via the 
two-gluon ($gg$) and $q\bar q$ $t$-channel exchange to be~\cite{Martin:2005hd}
\begin{equation}
  \frac{\left.({\rm d}\sigma_T^{\gamma^*p}/{\rm d}E_T)\right|_{\Pom=G}}
       {\left.({\rm d}\sigma_T^{\gamma^*p}/{\rm d}E_T)\right|_{\Pom=S}}\ =\ 
       \frac{81}{4}\left[\frac{\alpha(1-\alpha)Q^2}{E^2_T+\alpha(1-\alpha)Q^2}
	 \,
	 \frac{R_g\,x_{\Pom}g(x_{\Pom},\mu^2)}{R_q\,x_{\Pom}S(x_{\Pom},\mu^2)}
	 \right]^2,
\end{equation}
where $\alpha$ is the photon's light-cone momentum fraction carried by
the high-$E_T$ jet, and the scale $\mu^2=\alpha(1-\alpha)Q^2+E_T^2$.  
Therefore, measurements of exclusive diffractive dijets at large $E_T$
in photoproduction 
would probe the presence of the $q\bar{q}$ $t$-channel exchange.  However,
it remains to be 
seen whether a measurement of exclusive diffractive dijets is feasible 
without imposing a cut on $\beta$.

In summary, we have shown how to obtain {\it universal} diffractive
parton densities $a^{\rm D}$ which can be used in the description of
different diffractive processes.  We emphasise that the perturbative
QCD contribution originated by the inhomogeneous (last) term in
\eqref{eq:adevolve} is not small and, starting from a relatively low scale 
$\mu_0=1\,$GeV, may
generate up to a third of the diffractive parton densities.  Simultaneously, 
an account of this inhomogeneous term, and the twist-four 
$F_L^{{\rm D}(3)}$ contribution, leads to a lower
diffractive gluon density at high $z$ in comparison to that obtained using
the Regge factorisation hypothesis.  The presence of the perturbative (large 
scale) contributions reveals
itself in larger transverse momenta (with a power-like tail)
of the secondaries observed in the `Pomeron fragmentation' region.  The 
colour singlet sea-quark pair exchange is an important
component of the perturbative Pomeron, which plays the dominant role
in exclusive diffractive dijet production with $E^2_T\gg Q^2$.  
Parameterisations for the DPDFs and the diffractive structure functions 
are made publically available.\footnote{\texttt{http://durpdg.dur.ac.uk/hepdata/mrw.html}}

\section*{Acknowledgments}

M.G.R.~gratefully acknowledges the financial support of DESY and the II 
Institute of Theoretical Physics, University of Hamburg.  
G.W.~acknowledges the UK Science and Technology Facilities Council 
for the award of a Responsive Research Associate position.

\begin{footnotesize}
{\raggedright

}
\end{footnotesize}

\end{document}